\documentclass[11pt]{article}
\usepackage[margin=1in]{geometry}
\usepackage{graphicx}
\usepackage{booktabs}
\usepackage{xcolor}
\usepackage{tikz}
\usetikzlibrary{arrows.meta, positioning, calc}
\usepackage[colorlinks=true, linkcolor=blue!50!black, citecolor=blue!50!black,
            urlcolor=blue!50!black]{hyperref}
\usepackage{authblk}
\usepackage{microtype}

\newcommand{\refModules}{2}
\newcommand{\refChannelsPerModule}{16}
\newcommand{\refRuns}{30}
\newcommand{\refWfPerChannel}{400}
\newcommand{\refSamples}{1024}
\newcommand{\refPlanted}{7}

\newcommand{\ifContam}{0.07}
\newcommand{\ewHorizon}{5}
\newcommand{\refChannelRuns}{960}
\newcommand{\refFlagged}{89}

\newcommand{\refFlaggedChannels}{6}

\newcommand{\injRows}{2800}
\newcommand{\injBad}{2720}
\newcommand{\injClean}{80}
\newcommand{\ifAUC}{0.896}
\newcommand{\effML}{53.9}
\newcommand{\fprML}{0.0}

\newcommand{\effCuts}{33.2}
\newcommand{\fprCuts}{0.0}
\newcommand{\precCuts}{100.0}

\newcommand{\aucSameMean}{0.917}
\newcommand{\aucCrossMean}{0.888}

\newcommand{\fcSeries}{96}
\newcommand{\fcOffTrend}{3}
\newcommand{\fcEarlyWarn}{1}
\newcommand{\btPoints}{380}
\newcommand{\btCoverage}{0.96}

\newcommand{\nAlerts}{7}

\newcommand{\demoObjects}{9}
\newcommand{\demoLeaves}{6}
\newcommand{\demoCompared}{9}
\newcommand{\demoRootA}{demoDQM}
\newcommand{\demoRootB}{Monitoring/demoDQM}

\newcommand{\hcFprChi}{0.07}
\newcommand{\hcFprKs}{0.07}
\newcommand{\hcNegatives}{1500}
\newcommand{\hcRuns}{12}
\newcommand{\hcPairs}{36}
\newcommand{\hcPlanted}{3}

\title{\textbf{ARGUS: an experiment-agnostic framework for detector-health
monitoring in particle physics}}
\author[1]{Kamal Benslama}
\affil[1]{Drew University, Madison, New Jersey, USA}
\date{\today}

\begin{document}
\maketitle

\begin{abstract}
Every particle-physics experiment builds detector-health monitoring, and
many end up rebuilding it, because a first system written under
commissioning pressure often proves too rigid to maintain. ARGUS (Automated Anomaly-detection, Run-quality
and General Unified Surveillance) is a configuration-driven framework that
separates what is generic in this problem (the health monitors, their
validation, and the reporting) from what is not (the data schema and the
thresholds). Standing it up for a new detector requires a data adapter
and a configuration file, not a rewrite. Detector health is judged by four
independent monitors: expert cuts, which are authoritative; a
machine-learning second opinion that is additive and off by default; a
trend forecast that gives early warning before a threshold is crossed; and
a reference-histogram comparison with pluggable statistical tests. The
monitors are compared and never merged, and an alerting layer turns their
flags and warnings into deduplicated notifications. We demonstrate the
full chain on a
synthetic reference detector with planted pathologies and a fault-injection
campaign that manufactures ground truth: the cuts recover every planted
pathology with no false positives, the machine-learning monitor reaches a
ROC AUC of \ifAUC{} on injected faults and extends coverage to fault classes
no cut was written for, the forecast flags a drifting channel before its
cut fires, and the histogram comparison catches every planted shape
distortion while holding its configured false-positive rate on clean
pairs. A second demonstration shows the framework adapting to a
reorganized file layout with no change to code or configuration. Every
result and figure in this paper regenerates from the code alone. The
framework has been implemented for two particle-physics
experiments, where it is under evaluation; adoption decisions rest with
the collaborations.
\end{abstract}

% ============================================================================
\section{Introduction}
\label{sec:intro}

A running experiment needs to know, run by run and channel by channel,
whether its detector is healthy. The question sounds simple, but
answering it in practice is not: thousands of channels drift, die,
saturate, and go noisy in ways
that evolve over months, the monitoring code that watches them is usually
written under commissioning pressure, and the thresholds that define
``healthy'' are revised for years. Data-quality monitoring (DQM) systems
exist in every collaboration, and most share two structural weaknesses.
First, they are tied to one experiment: the file layout, the channel
numbering, the thresholds, and often the institution-specific
infrastructure are hardcoded, so none of the engineering effort transfers.
Second, their verdicts are rarely validated quantitatively: a monitoring
system that flags channels is trusted because experts wrote it, not because
its efficiency and false-positive rate have been measured against known
faults.

ARGUS approaches the problem from the opposite direction. It treats
detector-health monitoring as a generic analysis with experiment-specific
edges, and it treats the monitors themselves as objects of measurement. The
framework's engines, its validation harness, and its reporting layer contain
no experiment knowledge; everything an experiment must specify lives in a
data adapter and a single configuration file. The monitors' performance is
established the way a physics analysis would establish it: against
manufactured ground truth, with efficiencies, false-positive rates, and
receiver operating characteristics that anyone can regenerate.

This paper makes four contributions. It states a set of design principles
for trustworthy monitoring, of which the central one is that independent
health opinions must be compared and never merged
(section~\ref{sec:principles}). It describes an architecture that isolates
every experiment-specific assumption in one adapter and one configuration
file (section~\ref{sec:architecture}), with four monitors built on that
base (section~\ref{sec:monitors}). It introduces a fault-injection
validation method that measures the monitors quantitatively
(section~\ref{sec:validation}). And it demonstrates the entire chain on a
synthetic reference detector from which every result in this paper is
regenerated by one command (section~\ref{sec:reference}), including a
second demonstration in which the input file layout is reorganized and the
framework adapts with no change to code or configuration. Sections
\ref{sec:porting} through \ref{sec:outlook} cover the porting model,
related work, and planned extensions.

\section{Design principles}
\label{sec:principles}

\subsection{Independent opinions, compared and never merged}
Detector health in ARGUS is judged by parallel, independent monitors: expert
cuts, a machine-learning layer, a trend forecast, and a reference-histogram
comparison. They are shown side by
side and never averaged into a single verdict, because a disagreement
between monitors is itself the interesting signal. A channel the cuts pass
but the machine-learning layer flags deserves a look; a channel both flag
is almost certainly bad. The framework builds a worklist directly from these
disagreements, sorted so that contested channels surface first.

Within this structure the hierarchy is explicit. The cuts defined by the
detector experts are authoritative. The machine-learning layer is a second
opinion: it is additive, it is off by default, and when it is off the cuts
pipeline is byte-identical to a system with no machine learning in it. It
never overrides a cut verdict. This governance is what makes the layer
adoptable in practice: experts can turn it on, watch it, and turn it off,
without it ever having silently changed an answer they rely on.

\subsection{Monitoring is not physics}
A monitoring system answers ``has this changed'' rather than ``what is the
absolute value.'' The distinction has a practical consequence that is easy
to violate: DQM typically processes a sampled or prescaled fraction of the
data, so absolute counts and yields vary with the sampled fraction and must
never be trended raw across runs. Only intensive quantities (gains,
resolutions, fractions, ratios) or exposure-normalized rates (yield per
sampled event count or beam intensity) are comparable run to run. ARGUS
enforces this at the design level: trended quantities are intensive or
normalized, and the normalization machinery is part of the framework rather
than left to each user.

\subsection{Everything is configuration; structure comes from the data}
No cut, threshold, window, feature list, path, or label is hardcoded in the
engines. Every tunable lives in one configuration file with documented
defaults, which is what makes the same engines serve different detectors.

Structure is treated differently from semantics. The structure of the input,
meaning which detectors are present, where their objects live, and how
boards and channels are laid out, is discovered from the data itself,
because in practice file layouts churn as the upstream code evolves.
Semantics, meaning what a histogram represents and what thresholds apply,
follow an explicit precedence: a curated configuration overlay outranks
metadata embedded in the data, which outranks information extracted from the
upstream source code, which outranks generic defaults. When two of these
disagree, the disagreement is reported as drift and never silently applied.

\subsection{No hand-typed number}
Every figure, table, and quoted value on the monitoring website and in the
papers regenerates from the live results. A plot catalogue routes each plot
to the site and the documents, and a macro generator recomputes every quoted
number on each run. This paper practices the rule it states: every
measured value it quotes is a generated macro.

\section{Architecture}
\label{sec:architecture}

\begin{figure}[tb]
\centering
\begin{tikzpicture}[
  font=\small,
  node distance=4.5mm and 4mm,
  box/.style={draw=black!55, rounded corners=2pt, fill=black!4,
              align=center, inner sep=4pt, minimum height=7mm},
  branch/.style={box, fill=blue!8},
  side/.style={box, fill=orange!10, font=\footnotesize},
  arr/.style={-{Stealth[length=2.2mm]}, black!65, thick},
  dasharr/.style={-{Stealth[length=2mm]}, black!45, dashed}]

\node[box, text width=32mm] (raw) {raw waveforms\\\footnotesize per-run files};
\node[box, right=10mm of raw, text width=32mm] (hist)
  {DQM histograms\\\footnotesize per-run files};
\path (raw.south) -- (hist.south) coordinate[midway] (mid);
\node[box, below=4.5mm of mid, text width=86mm] (fe)
  {front end: discover $\cdot$ identify run $\cdot$ group $\cdot$
   combine (type-aware) $\cdot$ track};
\node[box, below=of fe, text width=58mm] (ad)
  {data adapter\\\footnotesize the only schema-aware component};
\node[side, left=6mm of ad, text width=27mm] (disc)
  {detector discovery\\source oracle\\(drift reported)};
\node[box, below=of ad, text width=52mm] (tab)
  {per-(channel, run) quality table};
\node[side, left=6mm of tab, text width=27mm] (enr)
  {exposure norm.\\conditions (IOV)};
\node[branch, below=8mm of tab, xshift=-56mm, text width=19mm] (a)
  {A $\cdot$ cuts\\\footnotesize authoritative};
\node[branch, right=3mm of a, text width=19mm] (b)
  {B $\cdot$ ML\\\footnotesize additive,\\\footnotesize off by default};
\node[branch, right=3mm of b, text width=19mm] (c)
  {C $\cdot$ forecast\\\footnotesize early warning};
\node[branch, right=3mm of c, text width=20mm] (d)
  {D $\cdot$ ref.\ compare\\\footnotesize per-hist tests};
\node[branch, right=3mm of d, text width=20mm] (v)
  {validation\\\footnotesize fault injection};
\path (c.south) -- (d.south) coordinate[midway] (bmid);
\node[box, below=8mm of bmid, text width=30mm] (cat) {plot catalogue};
\node[side, below=8mm of a.south, text width=24mm] (al)
  {alerting\\dedup $\cdot$ sinks};
\node[box, below=4.5mm of cat, xshift=-20mm, text width=42mm] (web)
  {website\\\footnotesize dashboards $\cdot$ runs $\cdot$ trends
   $\cdot$ search $\cdot$ worklist};
\node[box, right=6mm of web, text width=32mm] (pap)
  {papers\\\footnotesize live-number macros};
\node[side, anchor=south west, text width=20mm, minimum height=28mm] (cfg)
  at ($(v.north east)+(5mm,4mm)$)
  {one configuration file\\[1mm]thresholds\\windows\\features\\geometry\\branding};

\draw[arr] (raw) -- (fe.north -| raw);
\draw[arr] (hist) -- (fe.north -| hist);
\draw[arr] (fe) -- (ad);
\draw[arr] (disc) -- (ad);
\draw[arr] (ad) -- (tab);
\draw[arr] (enr) -- (tab);
\draw[arr] (tab.south) -- (a.north);
\draw[arr] (tab.south) -- (b.north);
\draw[arr] (tab.south) -- (c.north);
\draw[arr] (ad.south east) to[out=-60, in=105] (d.north);
\draw[arr] (tab.south) -- (v.north);
\draw[arr] (a.south) -- (cat.west);
\draw[arr] (b.south) -- (cat.north -| b);
\draw[arr] (c.south) -- (cat.north -| c);
\draw[arr] (d.south) -- (cat.north -| d);
\draw[arr] (v.south) -- (cat.east);
\draw[arr] (a.south) -- (al.north);
\draw[arr] (c.south) -- (al.north east);
\draw[arr] (cat.south) -- (web.north -| web);
\draw[arr] (cat.south) -- (pap.north -| pap);
\draw[dasharr] (cfg.west) -- (ad.east);
\draw[dasharr] (cfg.west) -- (tab.east);
\draw[dasharr] (cfg.south) -- (v.north east);
\end{tikzpicture}
\caption{The ARGUS architecture. Solid arrows are data flow; dashed arrows
mark the single configuration file that carries every tunable. The data
adapter is the only component that knows the input schema; the monitors,
the validation harness, the alerting layer (fed by the cut flags and the
forecast warnings), and the reporting layer are experiment-agnostic.}
\label{fig:architecture}
\end{figure}
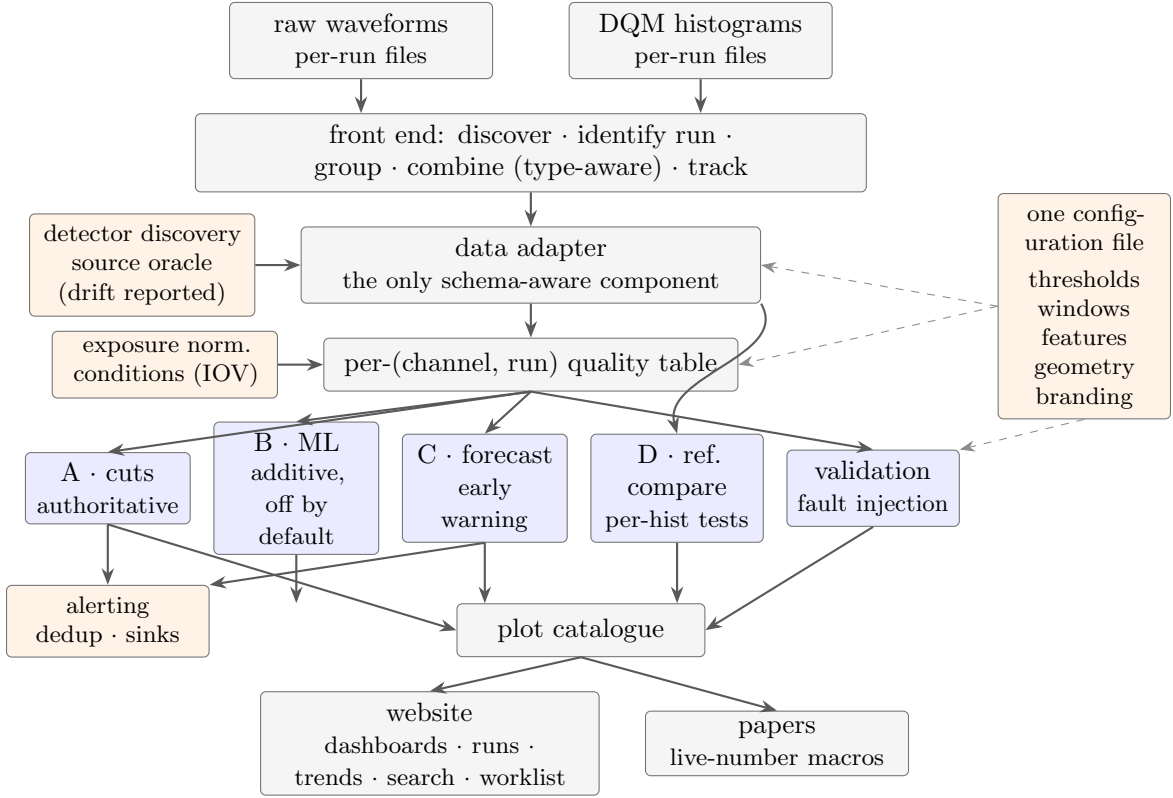

The framework is layered so that exactly one component knows the input
schema (figure~\ref{fig:architecture}). Per-run input files enter through
an adapter that emits a standard per-(channel, run) quality table; the
monitors, the validation harness, and the reporting layer all
operate on that table or on the adapter's standardized reading interface,
and none of them contain experiment knowledge.

\subsection{Ingest: two modes}
The adapter operates in one of two modes. In the raw mode, the input is
per-run files of digitized waveforms; the adapter extracts per-channel
features (baseline, noise, amplitude, signal-to-noise ratio, charge, timing
and shape statistics) itself. In the histogram mode, the input is the
monitoring histograms the experiment's own DQM jobs already produce; the
adapter reads them as they are. The second mode reflects a deliberate
stance: the framework reads what the experiment already produces and adds
its analysis on top, so no subsystem is asked to produce anything new. Both modes feed the identical downstream
table, and every monitor that operates on the table is independent of the
mode. The one exception is the waveform tier of the machine-learning
branch, which by construction needs the raw mode
(section~\ref{sec:branchb}); its channel-level tier runs in either mode.

\subsection{Front-end pipeline}
Ahead of the adapter sits a small pipeline that turns a directory of files
into well-defined run units: discover the files (directory, glob, or URL),
identify the run number (from the filename, from inside the file, or a
configured fallback), group files belonging to one run, and combine them.
Combination is type-aware, because a run interrupted and resumed produces
several files whose contents must not be merged naively: counts and
occupancies are summed, averaged profiles are combined with weights,
within-run trend histograms are concatenated in time order, and normalized
quantities are recomputed from the summed raw inputs rather than averaged.
A tracking manifest makes the pipeline incremental, so a watch mode can
analyze and publish new runs as they appear.

\subsection{Detector discovery and the source oracle}
The discovery engine locates each configured detector's subtree in the
input file from signature object names alone, with no folder paths
configured, and indexes every object under it by leaf name, type, and
embedded metadata. A detector configured in generic mode is rendered
entirely from that self-description, so extending monitoring to a new
subsystem is a configuration entry rather than new code.
Section~\ref{sec:demo-discovery} demonstrates this on a deliberately
reorganized file layout.

Semantics can additionally be cross-checked against the experiment's own
upstream monitoring source code, used as an advisory oracle: label orders,
enumeration definitions, and threshold constants are extracted defensively
and compared with the configuration. Agreement enriches the labels;
disagreement or absence produces a drift report and a fallback to the
configured values. The oracle can inform the analysis, and can never block
it.

\subsection{Reporting layer}
The output is an automatically rebuilt website driven by the same plot
catalogue that feeds the papers. Every plot carries its title, what it
shows, its axes, and a how-to-read note. The site provides per-detector
dashboards for the latest run, per-run detail pages, long-term trend pages,
a client-side comparison of any two runs, a site-wide search over plots and
definitions, and a worklist of channels on which the monitors disagree.
Views can be curated by audience, with a default for shifters (the
non-expert crew on monitoring duty) that hides
expert-only detail without restricting access to it. Summary tables are
downloadable, and a traffic-light rollup states each detector's status per
run.

\subsection{Enrichment inputs and per-detector extensions}
Three further inputs extend the core without changing it. Exposure
normalization joins sampled beam-intensity or event-count information so
that trended rates satisfy the comparability rule of
section~\ref{sec:principles}. Conditions ingest joins expert status
snapshots, matched by interval of validity, against the framework's
verdicts as an independent overlay, so the monitors can be compared with
the experiment's own bookkeeping. Finally, detector-specific analysis
modules can be plugged in beside the generic engines under the same
configuration and governance; a calibration-monitoring loop, in which
fitted calibration constants are extracted per run and then trended,
forecast, and validated like any other feature, is one deployed example.

\section{The four monitors}
\label{sec:monitors}

\subsection{Branch A: expert cuts}
The cuts branch computes per-(channel, run) flags: dead (the fraction of
good pulses falls below a threshold), noisy (the baseline noise is an
outlier), gain drift (the calibration constant departs from the channel's
own reference), uncalibrated (no valid calibration this run), and missing
(the channel disappeared from the readout). The noisy definition is
data-driven on two axes: a spatial rule flags a channel whose noise stands
out from its module in the same run (median plus a configured number of
median absolute deviations), and a temporal rule flags a channel whose
noise rises against its own history. A configurable absolute floor
suppresses relative flags in the quiet regime, and an optional hard ceiling
always fires. The gain reference is the channel's own history (a per-channel
median over a configurable window), so the drift flag needs no external
calibration database. Every flag records its provenance, meaning which rule
fired and with which values.

Where an experiment's upstream monitoring already computes official quality
verdicts, the branch splits into two sub-opinions: the official verdict is
ingested as canonical, and a derived cross-check recomputes flags from the
continuous quantities with configured thresholds that default to the
official ones. The two are displayed side by side and the official verdict
is never silently re-thresholded.

\subsection{Branch B: machine learning}
\label{sec:branchb}
The machine-learning branch has two tiers. The first tier, available in
the raw ingest mode, is a per-module
convolutional autoencoder trained on waveforms from a reference period; its
per-waveform reconstruction error is thresholded at a configured percentile
of the training distribution, and the fraction of a channel's waveforms
above that threshold becomes a per-(channel, run) feature, the abnormal
fraction. The second tier is a per-module Isolation Forest
\cite{isolation_forest} over the
channel-level features (the extracted statistics, the calibration gain and
its deviation, and the abnormal fraction); in the histogram ingest mode
the forest runs on the channel features alone. Channels missing a feature are
not discarded: indicator columns encode the missingness, which is itself
informative, since a channel with no calibration or no timing statistics is
telling the monitor something. Each flagged channel is reported with its
top contributing features, so an expert sees why the forest considered it
anomalous rather than a bare score. The expected anomalous fraction
(contamination) is the one prior the forest needs: it places the binary
decision threshold on the anomaly score, so it moves the operating point
but not the threshold-independent ROC AUC. Like every tunable it is
configuration, set per deployment by the user; the value used here,
\ifContam{}, is the framework default carried over unchanged, and it lands
near the reference detector's true planted anomaly rate without having
been tuned to it. The branch obeys the governance of
section~\ref{sec:principles}: additive, off by default, never overriding a
cut.

\subsection{Branch C: forecast}
The forecast branch watches trends. For each (channel, feature) series
across runs it maintains three quantities. The current level is an
exponentially weighted moving average: every past run contributes, but a
run's weight decays exponentially with its age, at a configurable
half-life, so the level tracks where the channel is now rather than where
it has been on average. The trend is a Theil--Sen
slope \cite{sen1968},\footnote{The Theil--Sen estimator is the median of the slopes of
the lines through every pair of points in the series. Because it takes a
median instead of minimizing squared residuals, a few anomalous runs
barely move it, where an ordinary least-squares fit would be dragged
toward them; up to about half the points can be outliers before it
breaks.} which is robust against the very anomalies the system exists to
find. The third quantity is a prediction band whose half-width is a
configured multiple of the median absolute deviation of the residuals
around the trend. A run outside the band is flagged off-trend. More useful
operationally, the branch projects each monitored series toward its limit,
which is reused from the cut thresholds rather than configured twice, and
raises an early warning when the projected crossing falls within a
configured horizon (\ewHorizon{} runs here). The branch runs standalone on
the quality table, reads no waveforms, and ships a backtest that scores its
own predictions on held-out runs, reporting band coverage and skill against
a persistence baseline.

\subsection{Branch D: reference-histogram comparison}
The fourth monitor is the classic DQM operation: compare each monitored
histogram against a reference and flag departures. Histograms are located
and paired between the run and the reference through the discovery engine,
by leaf name rather than by path, so the comparison survives layout
reorganization like everything else. Each pair is tested with a
per-histogram, configuration-selected statistical test; a chi-squared test
and a two-sample Kolmogorov--Smirnov test \cite{smirnov1948} ship with the
framework, and further tests plug into the same registry. The chi-squared compares shapes
with the reference scaled to the run's total, pools low-expectation bins
into one aggregate rather than dropping them (so a population appearing
where the reference is empty still registers), and takes each bin's
variance from the scaled reference, which keeps the test calibrated where
the naive observed-count variance does not. Verdicts are green, amber, or
red from configured p-value thresholds, and a pair with insufficient
statistics is reported as such rather than passed silently. The reference
is a designated golden run; because comparisons are by shape, prescaled
runs of different sizes compare cleanly. As with every branch, the
verdicts are one more independent opinion, and the branch is off by
default.

\subsection{The alerting layer}
A monitoring system that relies on people coming to look at it will,
sooner or later, not be looked at. ARGUS therefore does not only wait to
be read: it can call for attention. The alerting layer reads the
monitors' outputs and
raises a notification when a channel newly carries a standing flag (with a
configurable persistence requirement, so a one-run blip never alerts) or
when the forecast issues an early warning. Alerts are deduplicated against
a persistent state: a standing condition alerts once, clears when it goes
away, and alerts again on recurrence. Delivery is through configurable
sinks (a log file always; a webhook and email optionally, with the
endpoint taken from the environment rather than the configuration), and a
failing sink is reported and skipped, since alerting must never break the
analysis. Like every layer that is not the cuts, it is off by default.

\section{Validation by fault injection}
\label{sec:validation}

A monitoring system should be able to state its efficiency and its
false-positive rate. ARGUS measures both by manufacturing ground truth:
clean channels from held-out runs are corrupted at the level of the raw
waveforms with the eight fault classes of table~\ref{tab:faultmodel}. Each
fault is a parametrized transformation of the waveform, and its
\emph{severity} is the value of that parameter, in the fault's native
units; each class is injected over a configured grid of severities, several
of them deliberately too subtle to detect. Features are recomputed
with the same extractor the pipeline itself uses, so clean and corrupted rows are
internally consistent. The cuts re-evaluate each corrupted row using
reference statistics frozen from the real data, which is the faithful way
to ask whether the cuts would have caught this channel, since a fault does
not change its neighbors' baselines or its own history. The Isolation
Forest is trained on the untouched training period and applied to the
injected rows, which live entirely in the held-out period. The cuts and
the machine-learning monitor are scored against the same truth and are, as
everywhere in ARGUS, compared and not combined; the forecast has its own
backtest (section~\ref{sec:reference}) and the reference comparison its
own campaign (section~\ref{sec:hcvalidation}).

On \injRows{} injected-truth rows (\injBad{} faulty, \injClean{} clean
controls), the cuts reach an efficiency of \effCuts\% at a false-positive
rate of \fprCuts\% with a precision of \precCuts\%, and the
machine-learning monitor reaches \effML\% at \fprML\%, with an area under
the receiver operating characteristic curve (ROC AUC) of
\ifAUC{}\footnote{The ROC curve traces detection efficiency against
false-positive rate as the decision threshold on the anomaly score is
varied; the area under it summarizes the separation between faulty and
clean channels independently of any threshold choice. A perfect separator
has an AUC of 1, and random guessing gives 0.5.}
(figure~\ref{fig:roc}). The numbers say something precise about the
division of labor: the cuts catch what they were written for and
essentially nothing else, at almost no false-alarm cost, while the
machine-learning layer adds coverage on fault classes no cut was written
for. Table~\ref{tab:perfault} makes this explicit: afterpulsing, pulse
shifts, reduced light yield, and saturation are invisible to the cuts and
partially visible to the forest, while noise and dead channels are caught
by both.

The zero efficiencies in the cuts column are structural, not a tuning
accident. A rule-based cut inspects only the quantities its rule names,
and this cut set inspects the good-pulse fraction, the baseline noise, and
the calibration constant. Saturation clips the tops of the largest pulses
while leaving all three untouched, and a pulse-timing shift leaves them
untouched as well; the quantity a timing shift destroys is the charge
integral in the fixed integration window, which no cut inspects but the
forest does. A cut can therefore catch only the failure modes somebody
anticipated and wrote a rule for, and no severity changes that. The remedy on the cuts
side is straightforward and the framework supports it: an explicit
saturation rule, of the kind mature experiments carry among their official
quality metrics, is one configuration entry plus one rule, and the same
injection machinery then measures the new cut's efficiency the day it is
added. The remedy's limit is equally clear: there is always a next
unanticipated failure mode, and covering the unanticipated is what the
anomaly-detection branch is for.

The comparison also runs the other way, and the same table shows where the
cuts win. On gain shifts the cuts outperform the forest, because they read
the shifted calibration constant directly against the channel's own
reference while the forest must infer the shift from its indirect imprint
on the feature vector; a targeted rule with first-hand access to the
relevant quantity beats an untargeted model on that quantity's failure
mode. The cuts also hold a false-positive rate of \fprCuts\% with a
precision of \precCuts\%, an operating point a contamination-based anomaly
detector does not match uniformly. Neither monitor dominates the other,
which is the empirical justification for running both and comparing them
rather than merging them.

One limitation should be stated plainly. The efficiencies and
false-positive rates measured here are conditional on the injected fault
models: a real detector fails in ways nobody modeled, and no synthetic
campaign can bound the efficiency on those. The fault vocabulary is also
technology-specific: the classes of table~\ref{tab:faultmodel} model a
waveform readout, with its pulses, timings, and yields, and a tracking or
calorimetric system would define its own fault set for the same machinery
rather than reuse this one. What transfers is the method,
not the numbers: the injection machinery is part of the framework, so any
deployment can extend the fault set with its own observed failure modes
and remeasure both monitors against them, and the structural analysis
above (which quantities each monitor inspects) predicts where each will
and will not see.

Figure~\ref{fig:effsev} shows the per-fault turn-on curves, one
panel per fault class since severities are in each fault's native units;
the deliberately subtle severities (for example gain scale factors of 0.95
and 1.05) sit in the low-efficiency region of both monitors by design.
Figure~\ref{fig:sensitivity} shows the same information as a
detection-rate map over fault class and severity.

\begin{table}[tb]
  \centering
  \caption{The injected fault models and their severity parameters. The
  severity grids are configuration; the table is generated from it.}
  \label{tab:faultmodel}
  \small
  % AUTO-GENERATED fault-model table
\begin{tabular}{lp{0.40\textwidth}lp{0.18\textwidth}}
\toprule
fault & waveform-level model & severity & grid \\
\midrule
dead & post-baseline signal scaled by $(1-s)$ & suppression $s$ & 0.8, 1 \\
gauss noise & white noise added, amplitude $s$ times the channel's baseline RMS & noise amplitude $s$ & 1.2, 1.5, 2, 3, 5 \\
burst noise & coherent oscillation plus sparse spikes, amplitude $s$ times the baseline RMS & burst amplitude $s$ & 1, 2, 4, 8 \\
gain shift & pulse amplitudes and the calibration constant scaled by $g$ & scale factor $g$ & 0.95, 0.9, 0.7, 0.5, 1.05, 1.1, 1.3, 1.5 \\
reduced light & pulse amplitudes scaled by $a$, calibration constant unchanged & scale factor $a$ & 0.95, 0.9, 0.7, 0.5 \\
afterpulse & delayed copies of each pulse added with amplitude ratio $r$ & amplitude ratio $r$ & 0.1, 0.3, 0.6, 1 \\
saturation & pulses clipped at $(1-c)$ of their peak & clipped fraction $c$ & 0.05, 0.2, 0.5 \\
peak shift & pulse displaced by $d$ ticks, the vacated region refilled with baseline noise & shift $d$ (ticks) & 10, 20, 50, 100 \\
\bottomrule
\end{tabular}

\end{table}

\begin{figure}[tb]
  \centering
  \includegraphics[width=0.6\textwidth]{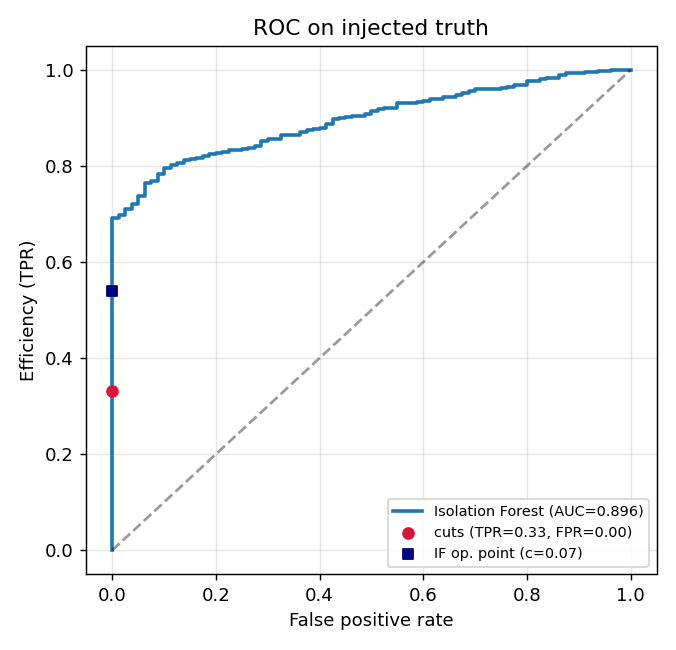}
  \caption{ROC of the Isolation-Forest score on injected truth, with the
  operating points of both monitors overlaid.}
  \label{fig:roc}
\end{figure}

\begin{figure}[tb]
  \centering
  \includegraphics[width=\textwidth]{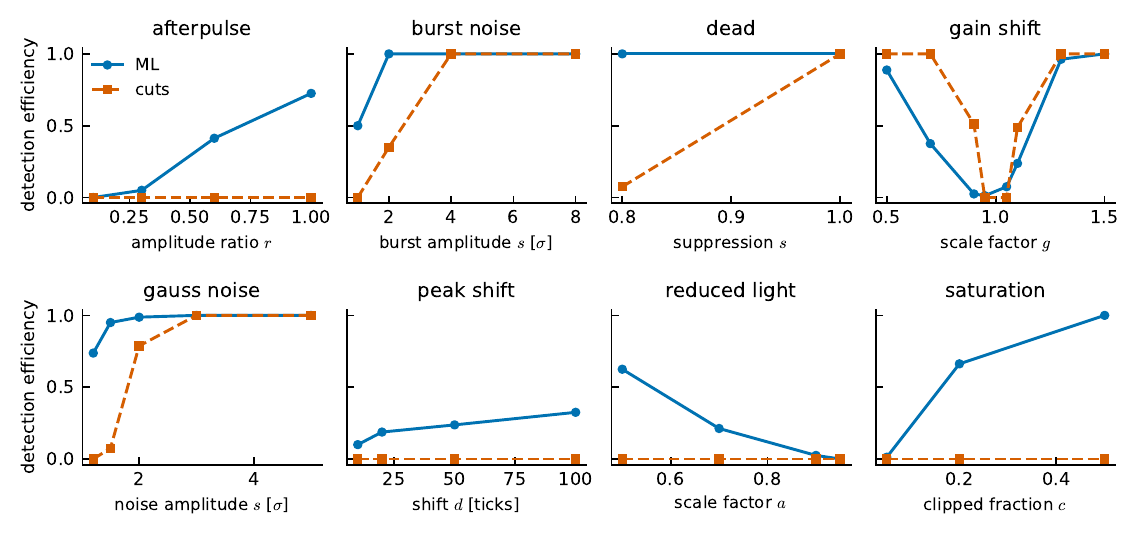}
  \caption{Detection efficiency versus fault severity, one panel per fault
  class (blue circles: machine learning, vermillion squares: cuts). Each
  panel's axis carries that fault's severity parameter from
  table~\ref{tab:faultmodel}, so the direction differs by panel: for the
  gain-shift and reduced-light scale factors, severity grows as the factor
  departs from one, while for the other faults larger values are more
  severe.}
  \label{fig:effsev}
\end{figure}

\begin{figure}[tb]
  \centering
  \includegraphics[width=0.9\textwidth]{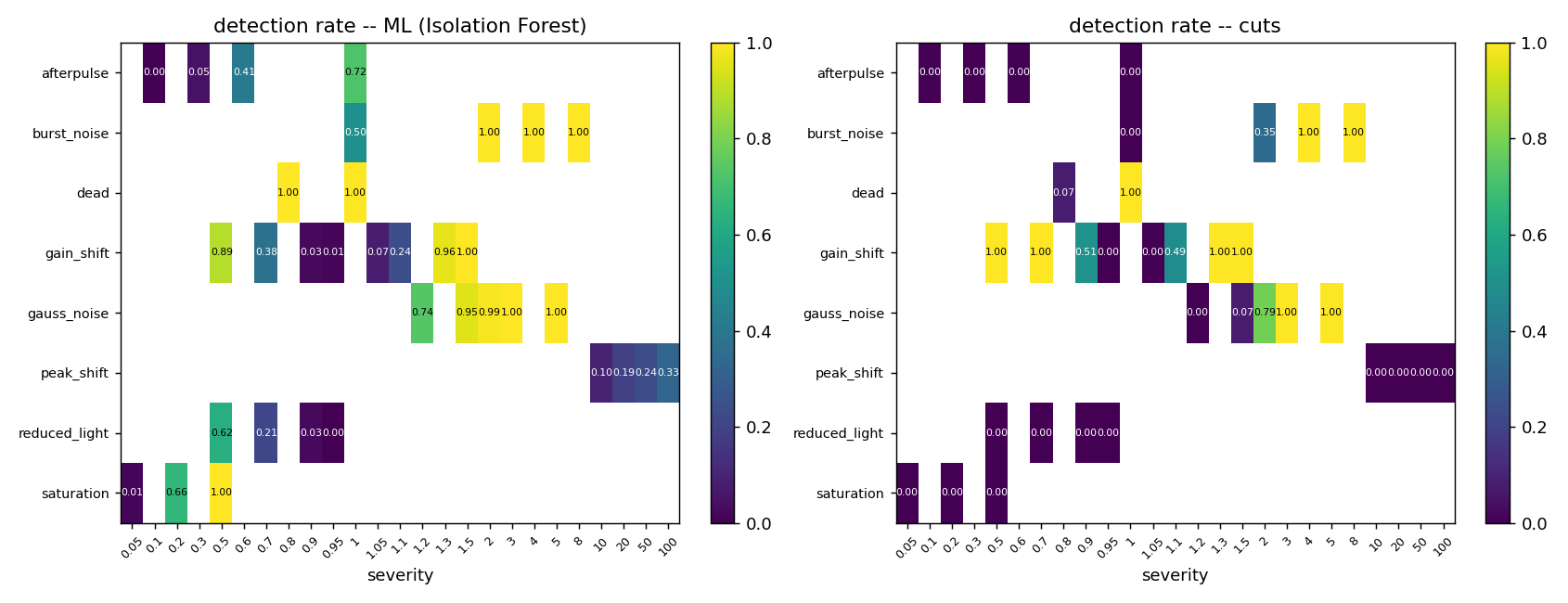}
  \caption{Detection rate per fault class and severity, for the
  machine-learning monitor (left) and the cuts (right).}
  \label{fig:sensitivity}
\end{figure}

\begin{table}[tb]
  \centering
  \caption{Detection efficiency by fault class on injected truth.}
  \label{tab:perfault}
  % AUTO-GENERATED per-fault efficiency table body
\begin{tabular}{lrrr}
\toprule
fault & injected rows & ML & cuts \\
\midrule
afterpulse & 320 & 30\% & 0\% \\
burst noise & 320 & 88\% & 59\% \\
dead & 160 & 100\% & 54\% \\
gain shift & 640 & 45\% & 62\% \\
gauss noise & 400 & 94\% & 57\% \\
peak shift & 320 & 21\% & 0\% \\
reduced light & 320 & 22\% & 0\% \\
saturation & 240 & 56\% & 0\% \\
\bottomrule
\end{tabular}

\end{table}

A transfer test probes whether the method depends on the detector
population it was tuned on: the forest is fitted on one module and applied
to the other module's injected rows (figure~\ref{fig:transfer}). The mean
ROC AUC is \aucSameMean{} when training and application module coincide and
\aucCrossMean{} when they differ, an essentially flat transfer, which is
the property that matters when the same configuration is carried to a
different detector. The binary operating points vary more between training
modules than the AUC does, since the flagged fraction follows the
configured contamination of whichever module the forest was fitted on;
the threshold-independent AUC is the transfer statement.

\begin{figure}[tb]
  \centering
  \includegraphics[width=0.85\textwidth]{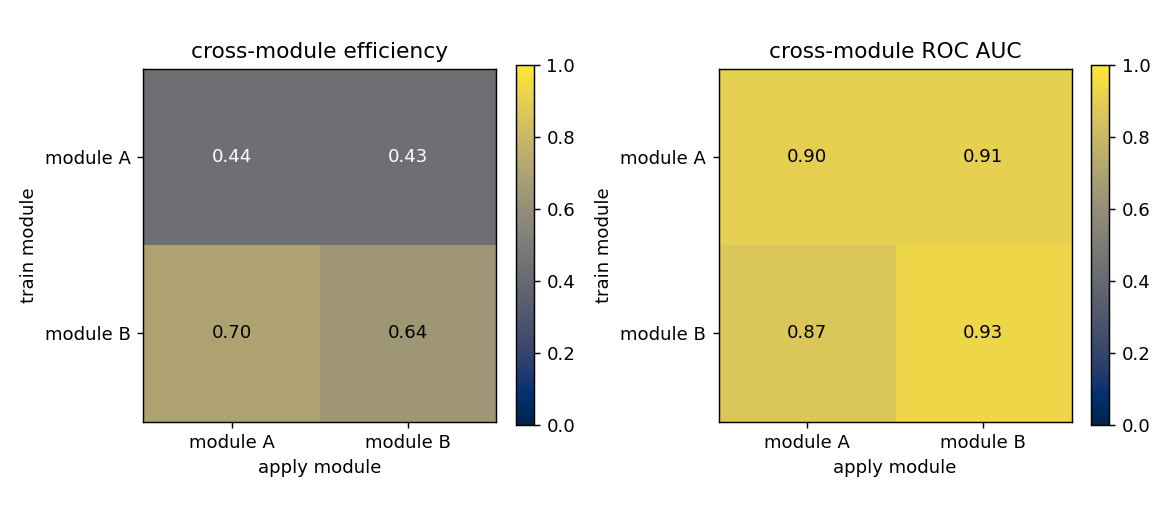}
  \caption{Cross-module transfer of the Isolation Forest on injected
  truth: detection efficiency (left) and ROC AUC (right) for each
  combination of training and application module.}
  \label{fig:transfer}
\end{figure}

\subsection{Validating the reference comparison}
\label{sec:hcvalidation}

The reference-histogram monitor is validated the same way, with
manufactured truth at the histogram level. Count-like spectra of
configured shapes are sampled, distorted with known truth (mean shifts,
width scaling, uniform tail contamination, and a thinned bin window) over
graded severities, and each shipped test is scored at the red threshold.
On \hcNegatives{} undistorted pairs the measured false-positive rate is
\hcFprChi\% for the chi-squared and \hcFprKs\% for the
Kolmogorov--Smirnov test, both consistent with the configured threshold,
which is itself a nontrivial statement for a binned chi-squared and the
reason the low-expectation pooling exists. Figure~\ref{fig:hceff} shows
the efficiency turn-on per distortion, and the two tests are visibly
complementary: the Kolmogorov--Smirnov test reaches full efficiency on
mean shifts as small as a tenth of the spectrum's width but is nearly
blind to symmetric tail growth, while the chi-squared dominates on
localized and tail effects. This complementarity is why the test is
selectable per histogram. A separate end-to-end check runs
\hcRuns{} synthetic monitoring runs through files and the discovery
pairing against a golden reference (\hcPairs{} comparisons): the
\hcPlanted{} planted distortions are all flagged red and no clean
histogram is.

\begin{figure}[tb]
  \centering
  \includegraphics[width=\textwidth]{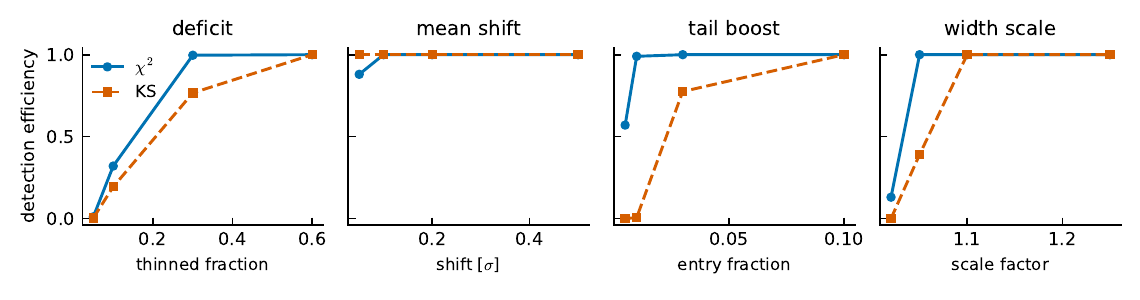}
  \caption{Detection efficiency of the reference-histogram comparison at
  the red threshold versus distortion severity, per distortion class, for
  the chi-squared (blue circles) and Kolmogorov--Smirnov (vermillion
  squares) tests.}
  \label{fig:hceff}
\end{figure}

\section{Reference-detector demonstration}
\label{sec:reference}

Everything quantitative in this paper comes from a synthetic reference
detector, generated and analyzed by one command. The generator writes
per-run files in the pipeline's exact input schema for a detector of
\refModules{} modules with \refChannelsPerModule{} channels each, across
\refRuns{} runs of \refWfPerChannel{} waveforms per channel
(\refSamples{} samples each), with per-channel pedestals, noise levels, and
single-photoelectron amplitudes drawn from seeded generators so the
detector has a stable identity, and calibration constants written
self-consistently with the pulse model. Synthetic data with known truth is
the stronger choice for validating monitors, not a concession: every
verdict can be checked against what was actually planted, and anyone with
the code can regenerate every figure without access to any experiment's
data. The entire chain, from data generation through every monitor,
validation, and figure, runs in well under an hour on an ordinary laptop
with no special hardware.

The generator plants \refPlanted{} pathologies: a noisy channel, a dead
channel, an uncalibrated channel, a channel that drops out of the readout,
a slow gain drift that crosses its threshold within the series, and two
slow risers that approach their limits only late in the series. The cuts recover the planted set and nothing else: \refFlagged{} of
\refChannelRuns{} channel-runs are flagged, every one of them on a planted
channel, with no false positives on clean channels.
\refFlaggedChannels{} of the \refPlanted{} planted channels carry flags;
the one that does not is the late gain drifter, which by design never
crosses a cut within the series. The forecast delivers the early-warning story on the rising-noise
channel: the projected crossing enters the warning horizon while the
channel is still passing every cut, and the noisy cut fires several runs
later, so the warning precedes the flag in the same series. This is the
operational value of the forecast branch in one measurement: the operator
learns about the channel before it fails rather than after. The late gain
drifter is flagged off-trend without having crossed any threshold. Of
\fcSeries{} forecast series, \fcOffTrend{} are off-trend and
\fcEarlyWarn{} carry an early warning; the backtest covers \btPoints{}
held-out points with a prediction-band coverage of \btCoverage{}. The
alerting layer, run over these results with the file sink, raises
\nAlerts{} notifications, one per standing planted condition plus the
early warning, and a second invocation raises none, which confirms the
deduplication works. On flat
channels the point forecast does not beat naive persistence, which is
expected, since a trend model only adds value when there is a trend;
the branch's operational value is the crossing projection, which
persistence cannot provide.

The framework has been implemented for two particle-physics
experiments, where it is being exercised and evaluated; whether it is
adopted is a decision that rests with each collaboration, and this paper
implies none. Everything shown here is computed from the synthetic
reference detector and regenerates from the code alone.

\subsection{Second demonstration: structure from the data}
\label{sec:demo-discovery}

The claim that structure is discovered rather than configured is tested
directly. A small synthetic detector (six object types over four boards)
is written into two files whose internal directory layouts differ
deliberately, as table~\ref{tab:layouts} shows: in layout A the objects
sit in flat groups under one top-level directory, while in layout B the
same objects are regrouped under renamed folders, the per-board
directories are renamed and renumbered, and the whole tree is nested one
level deeper. The directory names themselves (\texttt{demoDQM},
\texttt{Monitoring}, and the rest) carry no meaning to the framework;
they are arbitrary labels of the kind an evolving upstream code produces.
One configuration, which names no folder paths and identifies the
detector only by two signature histogram names, is applied to both files.
The discovery engine locates the detector subtree in each, at
\texttt{\demoRootA} in layout A and \texttt{\demoRootB} in layout B,
resolves the identical index of \demoObjects{} objects covering
\demoLeaves{} distinct histograms, and all \demoCompared{} resolved
objects agree bin for bin between the layouts. The layout change cost
zero code and zero configuration, which is the operational meaning of
structure from the data.

\begin{table}[tb]
  \centering
  \caption{Object locations in the two synthetic file layouts of the
  structure-from-data demonstration; $\langle b\rangle$ is the board
  number. The table is generated from the same configuration the
  demonstration ran with.}
  \label{tab:layouts}
  \small
  % AUTO-GENERATED by demo_discovery.py -- object locations per layout
\begin{tabular}{lll}
\toprule
object & layout A & layout B \\
\midrule
\texttt{h\_occupancy} & \texttt{demoDQM/All\_Histograms} & \texttt{Monitoring/demoDQM/Global/Occ} \\
\texttt{h\_baseline} & \texttt{demoDQM/All\_Histograms} & \texttt{Monitoring/demoDQM/Global/Noise} \\
\texttt{h\_rms} & \texttt{demoDQM/All\_Histograms} & \texttt{Monitoring/demoDQM/Global/Noise} \\
\texttt{h\_rate\_block} & \texttt{demoDQM/All\_Histograms} & \texttt{Monitoring/demoDQM/Trends} \\
\texttt{h\_hitmap} & \texttt{demoDQM/Summary\_Plots} & \texttt{Monitoring/demoDQM/Summary} \\
\texttt{ChannelQuality} & \texttt{demoDQM/Board$\langle b\rangle$/Histograms} & \texttt{Monitoring/demoDQM/Board\_$\langle b\rangle$} \\
\bottomrule
\end{tabular}

\end{table}

\section{Porting model}
\label{sec:porting}

The generic core comprises the four monitors, their validation machinery
(the fault injection, the forecast backtest, and the histogram-comparison
campaign), the discovery engine, the alerting layer, the catalogue-driven
reporting layer, and the statistics machinery behind the rule that no
number is hand-typed. The experiment-specific surface is deliberately small: a data
adapter (the input schema and, in raw mode, the channel-to-module mapping),
and one configuration file holding the geometry, the windows, the
thresholds, the feature lists, and the branding. Porting in practice means
writing the adapter, filling the configuration, and pointing the site
generator at a logo; the engines, the site, and the paper machinery carry
over unchanged. The reference detector of section~\ref{sec:reference} is
itself evidence of the model, since it is precisely a new ``experiment''
stood up by writing one generator and one configuration file against the
unchanged framework.

\section{Related work}
\label{sec:related}

Detector monitoring has an extensive literature, and ARGUS sits in a
specific place in it. The large collaborations operate mature DQM systems
that are deeply integrated with their experiments. The ATLAS data-quality
framework automatically checks tens of thousands of histograms per minute
against references with a library of predefined algorithms, feeding
operator displays and the offline quality bookkeeping
\cite{atlas_dqmf,atlas_dq_run2}. CMS complements its DQM with AutoDQM,
which applies beta-binomial tests, principal-component analysis, and
autoencoders to certified runs to rank anomalous ones
\cite{cms_autodqm}, and with an online autoencoder monitor for the
electromagnetic calorimeter \cite{cms_ecal_ae}; its Historic DQM trends
extracted quantities across long periods \cite{cms_hdqm}. LHCb's Monet
provides web-based monitoring with a configurable alarm system and
run-quality flagging \cite{lhcb_monet}, and the ALICE O2 quality control
runs user-defined checks synchronously with data taking inside the O2
system \cite{alice_o2qc}. At Jefferson Lab, Hydra applies computer vision
to monitoring plots and is deployed across the laboratory's experimental
halls \cite{jlab_hydra_first,jlab_hydra}. On the generic side, DQM4hep is
an event-model-agnostic C++ framework for online monitoring, used at
calorimeter testbeams \cite{dqm4hep}, and simulation-driven studies such
as MEDIC explore machine-learned data-quality methods on fast-simulated
detectors \cite{medic}.

These systems and ARGUS answer different parts of the same problem, and
table~\ref{tab:related} states the scopes side by side rather than ranking
them, since no head-to-head benchmark across experiments is possible. Two
things distinguish ARGUS's position. First, its generality lives at the
analysis and reporting layer with a configuration-only porting model,
demonstrated here by standing up a complete synthetic experiment against
the unchanged engines, whereas DQM4hep's generality lives at the online
transport layer and the collaboration systems are single-experiment by
design. For the same reason ARGUS is not tied to either side of the
online/offline divide: it consumes whatever monitoring products exist,
following data taking in watch mode with the latency of the upstream
monitoring job, and rerunning unchanged over offline or nearline
reprocessings. Second, its monitors ship with published, regenerable performance
measurements from injected ground truth, together with an explicit
governance rule for the machine-learning layer; this includes the classic
reference-histogram comparison that anchors the ATLAS and ALICE
frameworks, which ARGUS carries as its fourth monitor with a measured
false-positive rate and per-distortion efficiencies
(section~\ref{sec:hcvalidation}).

\begin{table}[tb]
  \centering
  \caption{Scope of related systems. The table states each system's focus;
  it is not a ranking, and no cross-experiment benchmark is implied.}
  \label{tab:related}
  \small
  \begin{tabular}{lllll}
    \toprule
    system & scope & layer & method focus & generality \\
    \midrule
    ATLAS DQMF \cite{atlas_dqmf} & ATLAS & online+offline & reference checks, thresholds & in-experiment \\
    CMS AutoDQM \cite{cms_autodqm} & CMS & offline & statistical tests, ML ranking & in-experiment \\
    CMS HDQM \cite{cms_hdqm} & CMS & offline & long-term trending & in-experiment \\
    Monet \cite{lhcb_monet} & LHCb & online+offline & web DQM, alarms, flagging & in-experiment \\
    ALICE O2 QC \cite{alice_o2qc} & ALICE & synchronous & user-defined checks & in-experiment \\
    Hydra \cite{jlab_hydra} & JLab halls & near-real-time & computer vision on plots & multi-hall \\
    DQM4hep \cite{dqm4hep} & testbeams & online & histogram transport/booking & EDM-agnostic \\
    MEDIC \cite{medic} & simulation & study & ML on fast simulation & not deployed \\
    ARGUS (this work) & any & online + offline & 4 monitors + validation & config-only port \\
    \bottomrule
  \end{tabular}
\end{table}

\section{Outlook}
\label{sec:outlook}

The plot catalogue's structured self-description, which already powers
the site-wide search, is the natural substrate for a conversational
interface over the monitoring results; the hooks exist behind a
configuration switch and are off by default, and coupling them to a
language-model endpoint is the next step. Finer time granularity is also
planned: the framework's interval abstraction already carries an optional
subrun beneath the run, so the same grouping machinery that combines
multi-file runs can analyze each subrun separately and roll the verdicts
up into a configured run-level verdict. The operational value is data
preservation: when a problem appears partway through a run, per-subrun
verdicts let an experiment discard only the affected subruns rather than
the entire run. A richer escalation policy for the alerting layer and a
public release of the framework core with the reference-detector
demonstration round out the plan.

\section{Summary}
\label{sec:summary}

ARGUS treats detector-health monitoring as a generic, measurable analysis.
Four independent monitors, with expert cuts authoritative, machine
learning additive and off by default, a forecast that warns before
thresholds are crossed, and a reference-histogram comparison with
pluggable tests, are compared and never merged, and an alerting layer
turns their verdicts into deduplicated notifications; every tunable is
configuration; structure is discovered from the data and semantic
disagreements are reported, never silently applied; and no number anywhere
in the system is hand-typed. On a fully synthetic reference detector, the
cuts recover every planted pathology with no false positives, the
machine-learning monitor reaches a ROC AUC of \ifAUC{} on injected faults
and extends coverage beyond the cuts, the forecast warns on a drifting
channel before its cut fires, the histogram comparison holds its
configured false-positive rate on clean pairs while catching every
planted shape distortion, and the framework adapts to a reorganized
input layout with no change to code or configuration. The framework has
been implemented for two particle-physics experiments, where it is under
evaluation, and everything in this paper regenerates from the code alone.

\section*{Code availability}

The framework is in place and available to experiments for use under
license from the author. A public release of the framework core together
with the reference-detector demonstration is in preparation; until it is
available, the code itself can be obtained from the author on request. The
released demonstration regenerates every result and figure in this paper
with one command.

\section*{Acknowledgments}

I thank my colleagues at Fermilab and CERN for many discussions on
detector monitoring, and Drew University for financial support. Both the framework's code and this manuscript were developed with
the assistance of an AI system (Anthropic), used as a programming and
drafting aid under my direction. I reviewed all code, results, and text
and bear sole responsibility for them; every quantitative claim in the
paper regenerates from the pipeline itself, independent of how any line of
it was first drafted.

\bibliographystyle{unsrt}
\bibliography{references}

@inproceedings{atlas_dqmf,
  author    = {Corso-Radu, A. and others},
  title     = {Data Quality Monitoring Framework for the {ATLAS} Experiment at the {LHC}},
  booktitle = {2007 IEEE Nuclear Science Symposium Conference Record},
  year      = {2007},
  doi       = {10.1109/NSSMIC.2007.4436303}
}

@article{atlas_dq_run2,
  author        = {{ATLAS Collaboration}},
  title         = {{ATLAS} data quality operations and performance for 2015--2018 data-taking},
  journal       = {JINST},
  volume        = {15},
  pages         = {P04003},
  year          = {2020},
  eprint        = {1911.04632},
  archivePrefix = {arXiv}
}

@article{cms_autodqm,
  note          = {arXiv:2501.13789, doi:10.1007/s41781-025-00147-2},
  author        = {{CMS Collaboration}},
  title         = {Anomaly Detection for Automated Data Quality Monitoring in the {CMS} Detector},
  year          = {2025},
  eprint        = {2501.13789},
  archivePrefix = {arXiv},
  doi           = {10.1007/s41781-025-00147-2}
}

@article{cms_ecal_ae,
  note          = {arXiv:2309.10157},
  author        = {{CMS ECAL Collaboration}},
  title         = {Autoencoder-based Anomaly Detection System for Online Data Quality Monitoring of the {CMS} Electromagnetic Calorimeter},
  year          = {2023},
  eprint        = {2309.10157},
  archivePrefix = {arXiv}
}

@article{cms_hdqm,
  author  = {Wightman, A. and others},
  title   = {A Historic Data Quality Monitor ({HDQM}) tool for the {CMS} Tracker Detector},
  journal = {EPJ Web Conf.},
  volume  = {214},
  pages   = {05030},
  year    = {2019},
  doi     = {10.1051/epjconf/201921405030}
}

@article{lhcb_monet,
  author  = {Anderlini, L. and others},
  title   = {{LHCb} data quality monitoring},
  journal = {J. Phys. Conf. Ser.},
  volume  = {898},
  pages   = {092027},
  year    = {2017},
  doi     = {10.1088/1742-6596/898/9/092027}
}

@article{jlab_hydra,
  author        = {Britton, T. and Jeske, T. and Lawrence, D. and Rajput, K.},
  title         = {Hydra: Computer Vision for Data Quality Monitoring},
  journal       = {JINST},
  volume        = {19},
  pages         = {C12005},
  year          = {2024},
  eprint        = {2403.00689},
  archivePrefix = {arXiv},
  doi           = {10.1088/1748-0221/19/12/C12005}
}

@article{jlab_hydra_first,
  note          = {arXiv:2105.07948},
  author        = {Britton, T. and Jeske, T. and Lawrence, D. and others},
  title         = {{AI} Enabled Data Quality Monitoring with {Hydra}},
  year          = {2021},
  eprint        = {2105.07948},
  archivePrefix = {arXiv}
}

@article{dqm4hep,
  author        = {Ete, R. and Pingault, A. and Mirabito, L.},
  title         = {{DQM4hep}: A generic data quality monitoring framework for {HEP}},
  journal       = {EPJ Web Conf.},
  volume        = {214},
  pages         = {05036},
  year          = {2019},
  eprint        = {1801.10414},
  archivePrefix = {arXiv},
  doi           = {10.1051/epjconf/201921405036}
}

@article{alice_o2qc,
  author  = {Konopka, P. and von Haller, B.},
  title   = {The {ALICE} {O2} data quality control system},
  journal = {EPJ Web Conf.},
  volume  = {245},
  pages   = {01027},
  year    = {2020},
  doi     = {10.1051/epjconf/202024501027}
}

@inproceedings{isolation_forest,
  author    = {Liu, F. T. and Ting, K. M. and Zhou, Z.-H.},
  title     = {Isolation Forest},
  booktitle = {2008 Eighth IEEE International Conference on Data Mining},
  pages     = {413--422},
  year      = {2008},
  doi       = {10.1109/ICDM.2008.17}
}

@article{sen1968,
  author  = {Sen, P. K.},
  title   = {Estimates of the regression coefficient based on {Kendall's} tau},
  journal = {J. Am. Stat. Assoc.},
  volume  = {63},
  pages   = {1379--1389},
  year    = {1968},
  doi     = {10.1080/01621459.1968.10480934}
}

@article{smirnov1948,
  author  = {Smirnov, N.},
  title   = {Table for estimating the goodness of fit of empirical distributions},
  journal = {Ann. Math. Stat.},
  volume  = {19},
  pages   = {279--281},
  year    = {1948},
  doi     = {10.1214/aoms/1177730256}
}

@article{medic,
  note          = {arXiv:2511.18172},
  author        = {Bassa, J. and Chattopadhyay, A. and Malik, S. and Escabi Rivera, M.},
  title         = {{MEDIC}: a network for monitoring data quality in collider experiments},
  year          = {2025},
  eprint        = {2511.18172},
  archivePrefix = {arXiv}
}
\end{document}